\documentclass[prb,preprint,showpacs]{revtex4}
\usepackage{graphicx}
\usepackage{dcolumn}
\usepackage{times}

\begin{document}

\title{The quasi-particle gap in a disordered boson Hubbard model
in two dimensions}

\author{Ji-Woo Lee$^1$ and Min-Chul Cha$^2$}
\affiliation {$^1$School of Physics, Korea Institute for Advanced Study, 
Dongdaemun-gu, Seoul 130-722, Korea \\
        $^{2}$Department of Applied Physics, Hanyang University, Ansan,
    Kyunggi-do 426-791, Korea}


\begin{abstract}
We investigate the behavior of the quasi-particle energy gap
near quantum phase transitions
in a two-dimensional disordered boson Hubbard model at a commensurate filling.
Via Monte Carlo simulations of ensembles with fixed numbers of particles,
we observe the behavior of the gap as a function of the tuning parameter
for various strength of diagonal disorder.
For weak disorder, we find that gapped Mott insulating phase is sustained
up to the transition point and disappears only in a superfluid,
strongly supporting a direct Mott-insulator-to-superfluid transition.
Bose glass behavior, insulating with vanishing gap, appears only
when the strength of disorder is bigger than a critical value.
\end{abstract}

\maketitle

\section{Introduction}
The interplay of interaction and disorder in quantum phase transitions
has been an important issue for past decades.
In systems of bosonic particles,
the phase coherence in a superfluid(SF) can be destroyed by
either interaction or disorder effect.
However, it is not obvious how these two effects interplay to drive
the quantum phase transitions.
To investigate the role of disorder,
the superfluid-insulator transitions of the disordered boson
Hubbard model\cite{Fisher89}
in low dimensions have been studied intensively
\cite{Scalettar91,Krauth91b,Singh92,Runge92,Wallin94,Pazmandi98,
Zhang95,Herbut00,Freericks96,Prokofev04}.
Experimental realization of this problem may include the transitions in
$^4$He in porous media\cite{Crowell97},
cold atoms in optical lattices\cite{Greiner02,Zoller03},
superconducting thin films\cite{Goldman98},
and small Josephson-junction arrays\cite{Fazio01}.

In the absence of disorder, a strong interaction suppresses the
density fluctuations and results in a Mott insulator (MI).
In the phase of MI it costs finite energy to add an extra particle.
This energy gap vanishes at the MI-SF transition
as the hopping strength increases,
allowing delocalized quasi-particles without any energy cost.
In the presence of disorder, the situation is subtle.
Based on a naive single-particle picture,
it has been argued
that an arbitrarily weak random potential will localize a single-particle
excitation as soon as the Mott gap vanishes.
Therefore, there exists an insulating phase with vanishing gap,
called the Bose glass (BG) phase,
which intervenes between MI and SF phases.
A hallmark of the BG-SF transition is that the value of the dynamical critical
exponent is given by $z=d$, where $d$ is the dimensionality of the system,
since the BG phase is compressible\cite{Fisher89}.

Many numerical searches
of the BG phase have been performed by identifying the
value of the critical exponents including $z$ in two dimensions.
Though the existence of BG for the strongly disordered case
has been generally accepted,
for the weakly disordered case it remains as a controversial problem.
Direct MI-SF transitions, without the intervening BG phase, have been found
\cite{Kisker97,Park99,Lee01,Lee04},
especially at commensurate fillings, 
while no clear numerical evidence of the BG phase has been reported in this case.
This may suggest that the physics of disordered system is divided into the
strong and the weak regimes, including the possibility that the particle-hole
excitations may screen out weak random potentials.
So far most of the numerical works focus on whether the critical exponents
support the direct MI-SF transition.
Therefore, it is very useful if one can directly observe the gap
near the superfluid-insulator transition to clarify whether there exists
the gapless insulating BG phase
between the MI and the SF phases in the presence of disorder.

In this work, we investigate the behavior of the quasi-particle gap
as a function of the tuning parameter for different strength of
disorder in a disordered two-dimensional boson Hubbard model. We
find that for weak disorder the gap vanishes sharp at the
transition, strongly supporting the existence of the direct MI-SF
transition. This feature disappears for strong disorder, thus
supporting the scenario that the disordered cases are divided into
the strong and the weak disorder regimes.

\section{Model}
Interacting bosons moving in the presence of a random potential
can be described by a boson Hubbard model
\begin{eqnarray}
{H}_{bH} =\frac{U}{2}\sum_j n_j^2 - \sum_j \mu_j n_j
-{t}\sum_{\langle i j \rangle} (b^\dagger_i b_j + b^\dagger_j b_i),
\end{eqnarray}
where $b_j(b_j^\dagger)$ is the boson annihilation (creation)
operator at $j$-th site, $n_j$ is the number operator, $U$ and $t$
are strength of the on-site interaction and the hopping amplitude
between nearest neighboring sites ${\langle i j \rangle}$,
respectively, and $\mu_j$ denotes a diagonal random potential due to
local impurities. We take a random potential $-\Delta <\mu_j <\Delta$,
where $\Delta$ is the strength of disorder, to make its average to
be zero for the case of a commensurate density. The phase diagram of
this model with possible three phases, known as the MI, BG, and 
SF, has drawn a great deal of attention, focusing on whether the BG
phase intervening between the MI and the SF phases exists even if the
disorder is weak.

It is believed that in the limit of large density
these phases and the transition between them
can be described by the quantum rotor model\cite{Sachdev}
\begin{eqnarray}
{H}_{\rm qr} =
{U\over 2}\sum_{j}(\frac{1}{i}\frac{\partial}{\partial \theta_j} -
{\bar \mu}_{j} )^2
- J\sum_{\langle i,j\rangle} \cos(\theta_i - \theta_j),
\end{eqnarray}
where $J=t n_0$ ($n_0$ is the average integer number of bosons per site),
${\bar \mu}_j=\mu_j/U$, and
$\theta_j$ is the phase angle of bosons at $j$-th site.

For numerical simulations, it is very useful to write the partition function
$Z=\sum_{\{\theta\}}\exp(-\beta H_{\rm qr})$ in terms of dual current
variables so that
\begin{eqnarray}
Z=\sum_{\{ \bf J\} }^{\nabla \cdot {\bf J}=0} e^{-H_{\rm cl}/K}
\end{eqnarray}
with the corresponding classical Hamiltonian\cite{Wallin94}
\begin{eqnarray}
H_{\rm cl}[{\bf J}] =  \sum_{(j,\tau)} \big\{\frac{1}{2}[J^{x^2}_{(j,\tau)}
+ J^{y^2}_{(j,\tau)} + J^{\tau^2}_{(j,\tau)}] - {\bar \mu}_j J^\tau_{(j,\tau)}
\big\},
\label{cl_hamil}
\end{eqnarray}
where $K \sim \sqrt{J/U}$ is the tuning parameter controlling the
quantum fluctuations. Here ${\nabla \cdot {\bf J}=0}$ means the
current conservation condition at each site $(j,\tau)$. Note that in this
representation the winding number in $\tau$-direction is the net
number of bosons, $N$. In other words, $N=0$ means a commensurate
density, and $N=1 (N=-1)$ represent the existence of extra
quasi-particle (hole), etc.

When temperature $T \to 0$, we expect $Z \approx e^{-\beta E_0}$, where $E_0$ is
the ground state energy of the quantum Hamiltonian.
Thus $E_0$ is closely related with the free energy of the classical action,
$F_{\rm cl}$.
This allows us to define a quasi-particle gap
\begin{eqnarray}
\Omega_g={1\over KL_\tau}\big\{F_{\rm cl}^{(N=1)}+F_{\rm cl}^{(N=-1)}
-2F_{\rm cl}^{(N=0)}\big\}
\end{eqnarray}
from the free energy with a fixed number $N$. For finite systems of
size $L\times L\times L_\tau$, where $L$ is the size in a spatial
direction and $L_\tau$ in the temporal direction, we use $L_\tau$ in
place of the inverse temperature $\beta=1/T$. Even though the free
energy, equivalently the gap, is hard to observe directly, its
derivative with respect to $1/K$
\begin{eqnarray}
-K^2\frac{d\Omega_g}{dK}=\frac{1}{L_\tau} \bigg\{ 
\langle H_{\rm cl}\rangle^{(N=1)} 
& \hspace{-1.0cm}+\langle H_{\rm cl}\rangle^{(N=-1)} & \nonumber
\\
&~~~-2\langle H_{\rm cl}\rangle^{(N=0)}\bigg\} &
\label{eq:gap_behavior}
\end{eqnarray}
is easily obtainable by Monte Carlo method adopting the classical
Hamiltonian in Eq.~(\ref{cl_hamil}).

\begin{figure}[t]
\includegraphics[width=8cm]{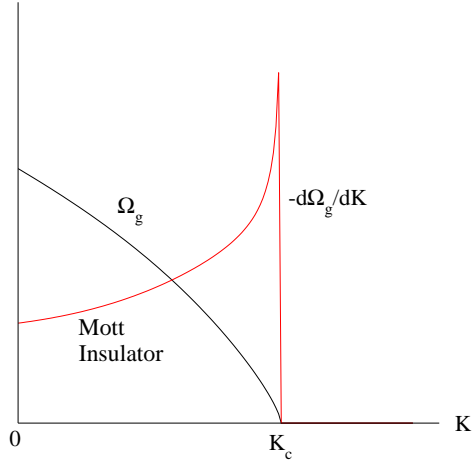}
\caption{\label{fig:gap} A schematic plot for the behavior of the
energy gap, $\Omega_g$, and its negative derivative with respect to
$K$, $-d\Omega_g/dK$. As gap vanishes continuously at the
transition, $-d\Omega_g/dK$ increases sharp.}
\end{figure}

Suppose the gap vanishes at a continuous transition with the dynamical
critical exponent $z$ and the correlation length critical exponent $\nu$.
Since we have $\Omega_g \sim |K-K_c|^{z\nu}$ as
we approach the transition below,
we expect a diverging behavior $-d\Omega_g/dK \sim |K-K_c|^{z\nu-1}$
when $z\nu <1$,
as schematically shown in Figure~\ref{fig:gap}.
However, if the gap vanishes before the correlation diverges,
the diverging behavior of $-d\Omega_g/dK$ would not appear.
By observing $d\Omega_g/dK$, therefore, especially its feature of the sharp
rise and the subsequent decline, we are able to 
investigate the behavior of the gap.

The finite-size scaling ansatz of the gap from the hyperscaling relations
can be constructed as
\begin{eqnarray}
\Omega_g=L^{-z}X_\Omega \big((K-K_c)L^{1/\nu},L_\tau/L^z\big),
\end{eqnarray}
where $X_\Omega$ is a scaling function.
By taking a derivative, we have
\begin{eqnarray}
-K^2\frac{d\Omega_g}{dK}=L^{1/\nu-z}X_\Omega^{\prime}
\big((K-K_c)L^{1/\nu},L_\tau/L^z\big)
\label{eq:scaling}
\end{eqnarray}
where $X_\Omega^{\prime} \equiv \partial X_\Omega/\partial (1/K)$ is
another scaling function. Since it is believed that the gap vanishes
for Bose glass phase and superfluid phase, we expect
$d\Omega_g/dK=0$ in those phases.

\section{Monte Carlo Method}

To investigate the behavior of the gap at a commensurate density,
the expectation value of the energy in the equivalent classical
model are calculated in three replica with $N=0,1$ and $-1$ for the
same complexions of the random potential $\bar{\mu}_j$, and, then,
average over different random potentials are taken. We adopt a worm
algorithm\cite{Alet03} to update the current configurations in
Eq.~(\ref{cl_hamil}) while rejecting any updates changing the the
winding numbers in $\tau$-direction in order to keep the number of
bosons fixed. In addition, to create a pair of particle and hole,
which also does not change the net winding number in
$\tau$-direction, a conventional method of the global loop-current
update\cite{Wallin94} is performed in $\tau$-direction.

\begin{figure}[t]
\includegraphics*[width=8cm]{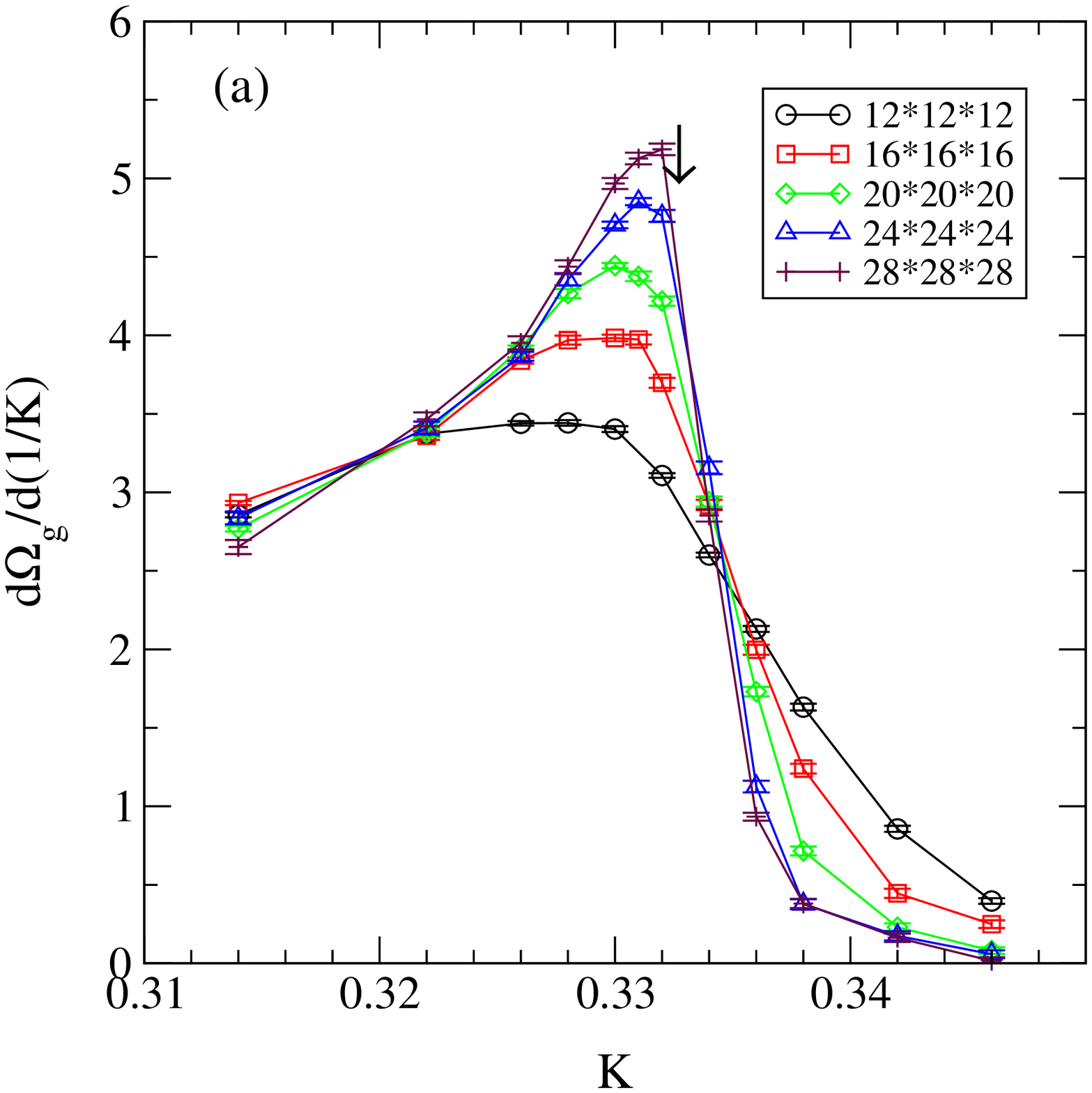}
\includegraphics*[width=8cm]{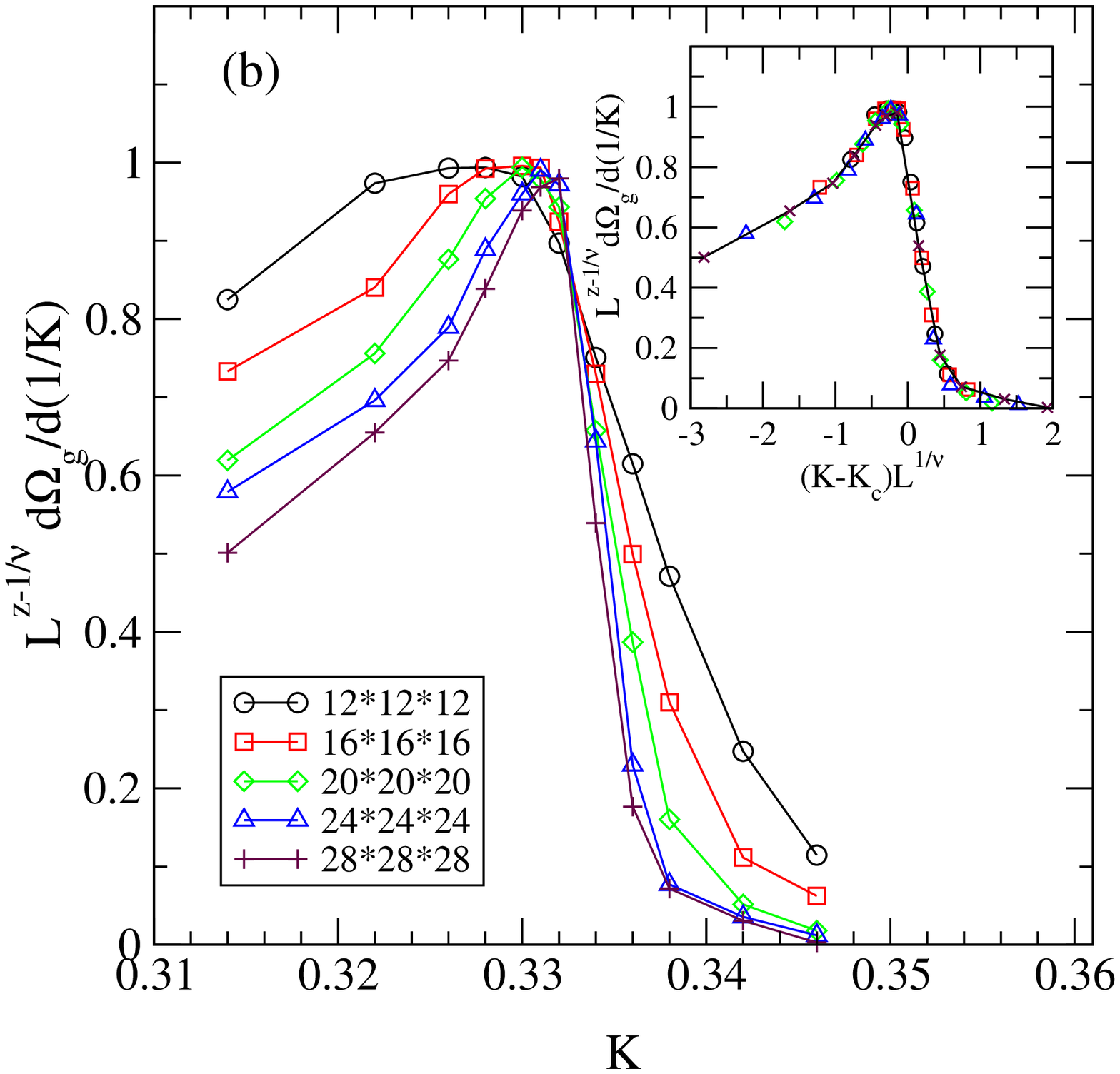}
\caption{
The behavior of the gap as a function of $K$ for the pure case.
(a) A sharp decline of $-d\Omega_g/dK$ is observed near the critical
point $K_c=0.333$, marked by an arrow, which is a signal of the vanishing gap.
(b) The properties of the gap shows a finite-size scaling behavior with
the dynamical critical exponent $z=1.0$ and the correlation length
critical exponent $\nu=0.67$. Inset shows this scaling behavior
as a function of the scaling variable $(K-K_c)L^{1/\nu}$.
High quality data collapsing onto a curve for different sizes
strongly suggests that the gap indeed follows a scaling relation.}
\label{fig:pure}
\end{figure}

Figure~\ref{fig:pure} shows the behavior of the gap as a function of
$K$ for pure case ($\Delta=0$). Near the critical point $K_c=0.333$\cite{Alet03}
marked by an arrow in the figure, a stiff rise and the subsequent
decline of $d\Omega_g/d(1/K)$ appears as shown in Fig.~2a. This
feature becomes sharper in bigger systems, indicating that the gap
indeed vanishes at the critical point. Interestingly, this behavior
satisfies the scaling ansatz of the gap, given in
Eq.~(\ref{eq:scaling}) as we can find in Fig.~2b. Here we just
confirm that the known values of the exponent $z=1.0$ and $\nu=0.67$
yield a high quality scaling behavior, such as the crossing behavior 
of $L^{z-1/\nu}d\Omega_g/d(1/K)$ at
$K=K_c$ for different sizes as a function of $K$ and data collapsing
as a function of $(K-K_c)L^{1/\nu}$.

\begin{figure}[t]
\includegraphics[width=8cm]{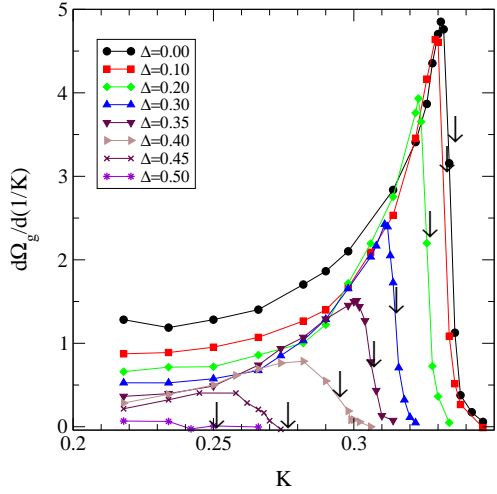}
\caption{
The behavior of the gap as a function of $K$ for different $\Delta$
in the systems of the size $L\times L\times L_\tau=24\times 24\times 24$.
The critical points for given value of $\Delta$, estimated from
the finite-size scaling behavior of the superfluid stiffness, are
marked by the arrows.
}
\label{fig:delta}
\end{figure}

The success in understanding the critical properties for the pure
case through the behavior of the gap motivates a extension of the
study to the disordered case. Figure~\ref{fig:delta} shows the
behavior of the gap as a function of $K$ for different $\Delta$. The
feature of the sharp rise and decline in $d\Omega_g/d(1/K)$ near the critical point
appears for small $\Delta$. However, this feature is absent for
$\Delta > \Delta_c \approx 0.4$. In this strong disordered regime,
we have $d\Omega_g/dK \approx 0$ near the transition point,
suggesting that only in this strong disorder regime BG phase
intervenes between the MI and the SF phases.
These results are well consistent with the previous studies\cite{Lee01,Lee04}
of this model through the finite-size scaling properties of
the superfluid stiffness and the correlation functions, showing 
the different critical exponents in the strongly and weakly disordered regimes
separated by the critical strength of disorder $\Delta_c \approx 0.4$.

\begin{figure}[h]
\includegraphics[width=8cm]{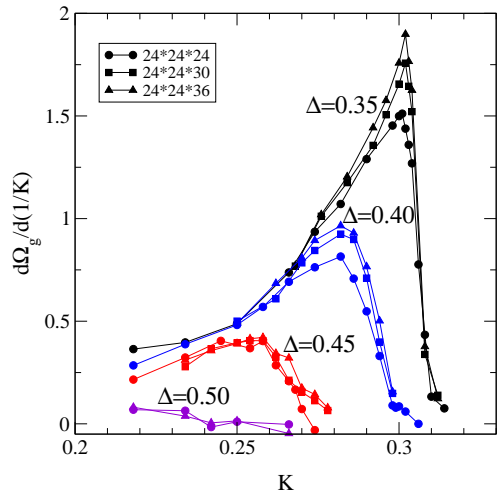}
\caption{
The behavior of the gap for different sizes near $\Delta_c$.
A qualitative change of the gap behavior appears
across $\Delta_c \approx 0.4$.
}
\label{fig:size}
\end{figure}

Figure~\ref{fig:size} shows the size dependence of the gap feature near
$\Delta_c \approx 0.4$.
For $\Delta < \Delta_c$, the feature of the sharp rise becomes apparent
in bigger systems. No such feature appears for $\Delta > \Delta_c$.
This implies that there is a qualitative change of the gap feature
across the critical strength of the disorder, $\Delta_c$.

\section{summary}
In summary, we have investigated the behavior of the quasi-particle gap
as a function of the tuning parameter $K$ for different strength of
disorder $\Delta$ in a two-dimensional boson Hubbard model,
focusing on whether the BG phase exists between the MI
and the SF phases in the presence of random potentials.
The gap properties are studied in an equivalent classical model
with a fixed number of particles.
In the pure case, the properties of the gap satisfies
a finite-size scaling relation with a sharp signal
indicating that the gap vanishes at the transition.
In the disordered case,
our results strongly support a scenario that the cases
are divided into the strong and the weak disordered regimes.
For weak disorder, we can find the features of the vanishing gap
at the transition, implying that the direct MI-SF transition occurs.
This feature does not appear in the strong disorder regime,
which supports the existence of the compressible BG phase in this case.

\acknowledgements This work was supported by Korea Research Fund
grant No. R05-2004-000-11004-0. MCC appreciates the hospitality of
the Korea Institute of Advanced Study.

\end{document}